\documentstyle[preprint,aps]{revtex}
\begin{document}
\draft
 
\title{Non-perturbative chiral approach to 
$K^- p \rightarrow \gamma Y$ reactions}
\author{                                    
T.-S. H. Lee$^{1,2}$,
J.A. Oller${^1}$,
E. Oset$^{1}$ and A. Ramos$^{3}$}

\address 
{$^1$Departmento de F\'{\i}sica Te\'orica and IFIC, Centro Mixto Universidad de
Valencia-CSIC, 46100 Burjassot (Valencia), Spain
\linebreak
$^2$ Physics Division, Argonne National Laboratory, Argonne, Illinois 60439,USA
\linebreak
$^3$ Department d'Estructura i Constituents de la Mat\`eria, Universitat de
Barcelona, Diagonal 647, 08028 Barcelona, Spain}
\date{\today}
 
\maketitle
\bigskip
\bigskip

\begin{abstract}
A recently developed nonperturbative chiral approach to
describe the $S=-1$ meson-baryon 
reactions has been extended to investigate the near threshold
$K^- p \rightarrow \gamma \Lambda, \gamma \Sigma^0$ reactions. 
With the parameters governed by chiral SU(3) symmetry, we show that
the predicted branching ratios 
$\Gamma_{K^- p \rightarrow \gamma \Lambda}/\Gamma_{K^- p \rightarrow all}$
and $\Gamma_{K^-p\rightarrow \gamma \Sigma^0}/\Gamma_{K^- p \rightarrow all}$ 
are close to the experimental
values. The coupling with the $\eta$ channels, which was shown to be 
important in the $S=-1$ meson-baryon reactions, is also found
to be significant here. Our results are consistent with the 
interpretation of the $\Lambda(1405)$ as a quasi-bound meson-baryon
state as found in other similar chiral approaches.
\end{abstract}

\bigskip
\bigskip
\bigskip


\bigskip
\bigskip
\newpage
\section{Introduction}

The near threshold $K^- p \rightarrow \gamma Y$ reaction with $Y=
\Lambda, \Sigma^0$ has long attracted a lot of interest, mainly because
of the possibility of using this reaction to resolve the 
debates\cite{jones,darewych,veit,zhong,kaxiras,he,arima} over
the structure of the $\Lambda(1405)$ resonance. Most of the earlier theoretical
investigations\cite{workman} neglected the initial strong $K^- p$ interactions.
It was first demonstrated by Siegel and Saghai\cite{saghai} 
that the initial $K^- p$ interactions can drastically change the predicted
capture rates and thus can significantly alter the interpretation of the data.
With the phenomenological separable 
potentials, they, however, needed an about $30 - 50 \%$ deviation 
of the coupling
constants from the SU(3) values to obtain an accurate description
of the data. 

Obviously, the progress can be made only when the initial strong interactions
and the photoproduction amplitudes are consistently described within the
same theoretical framework. For investigating the 
low energy processes involving strange
hadrons, such as the processes considered in this work, 
it is now generally believed
that the best starting point is the effective
chiral lagrangian with SU(3) symmetry. The most developed approach in
this direction is  Chiral Perturbation 
Theory\cite{gasser,meissner,pich,ecker,bernard}. 
Such an approach, however, becomes powerless in facing the $S=-1$ meson-baryon
interactions because of the formation of a resonance just below
the $K^- p$ threshold. An alternative nonperturbative chiral approach was
first developed by Kaiser, Siegel and Weise\cite{kaiser}. The essential idea
is to define the meson-baryon potentials by the SU(3) chiral lagrangian and
sum a subseries of the chiral expansion by solving a coupled-channel
Lippmann-Schwinger equation. They consider 
the lowest order and next to lowest order chiral lagrangians, and
implemented suitable form factors to make the resulting meson-baryon 
potentials of finite range such that the loops become convergent. 
The range parameters of form factors and some 
of the second order chiral lagrangians are fitted to the data. This
unitary coupled-channel approach generates dynamically 
the $\Lambda(1405)$ resonance as a quasibound meson 
baryon state, and reproduces the low energy data for  $K^-p$ elastic 
and inelastic channels. 
The generation of the $\Lambda$(1405) resonance is actually not a merit 
of the chiral lagrangians since it can be obtained in a suitable coupled
channel K matrix approach, which would implement unitarity in the
amplitudes \cite{saghai,dalitz,martin}. The merit of the chiral 
lagrangians is that they can provide an expansion for this K matrix consistent 
with chiral symmetry and its breaking in the different strangeness channels, 
$S=0$,1,$-$1. The extension of the work of \cite{kaiser} to the $S=$0,1 channels 
is done in Ref. \cite{kaiserbis}. Particularly, 
when working in the S=0 sector, $\pi N$ 
and coupled channels, the $N^*(1535)$ resonance 
is also generated dynamically. Simultaneously, the $\eta$ and K 
photoproduction processes in the S=0 channels were also studied in a similar 
approach, and with a few more parameters a global reproduction of the 
strong and electromagnetic cross sections was obtained. 

The work of \cite{oset} was inspired by Ref.\cite{kaiser}, but followed 
closely the approach developed in the study 
of the s-wave meson-meson scattering of
Ref.\cite{oller}. Starting with a coupled-channel equation based on the lowest
order chiral lagrangian, it was found that the off shell effects in 
the vertices can be absorbed in coupling constants renormalization. Hence, 
only the on-shell part of the vertices needed to be considered. The loop 
integrals were regularized by means of a cut off, rather than using a 
form factor as done in \cite{kaiser}, which allows one to keep the chiral 
logarithms. The resulting coupled-channel equation is similar to that of
Ref.\cite{kaiser}. Another major step taken in Ref.\cite{oset} was to 
include 
the $\eta$ and $\Xi$ channels in order to have exact SU(3) symmetry when
the mass differences among the members of each Octet, mesonic or baryonic, 
are neglected. 
Both channels are not opened at 
low energies and the loop integrations involving these two channels 
only contribute to the real part of the 
amplitudes. The inclusion of the $\Xi$ channels was found to have negligible 
effects, but the $\eta$ channels were found numerically 
important. The approach 
of \cite{oset} turns out to be more economical because with the use of 
only one cut off parameter and the input of the lowest order lagrangian 
one can reproduce fairly well all the low energy data in the $S=-1$ sector. 
However, the fact that the $\eta$ meson loops only provide a real part to the 
amplitudes, allows the effect of these channels to be effectively taken into 
account by means of the parameters of the second order lagrangians, as done 
in \cite{kaiser}, provided one is not too close to the thresholds of these 
channels. Conversely, the effects of the second order lagrangian 
can be effectively incorporated by means of a suitable cut off in \cite{oset}, 
much as it happened in the study of the scalar sector in the meson-meson 
interaction in \cite{oller,ramonet}. The physics contained in the approaches of 
\cite{kaiser,oset} is basically the same, in spite of the different 
inputs and treatment of the scattering equations, and the results are 
remarkably similar. 
However, it is not clear that this simple approach can be extended to the 
$S=0$ and $S=1$ sectors because higher order counterterms, which cannot be 
accounted for by means of just one cut off, could appear. Preliminary results 
for the meson-baryon interaction \cite{assum} also point in the same direction. 

In this paper, we extend the unitary coupled channel chiral approach of 
\cite{oset} in order to investigate the $K^- p \rightarrow \gamma \Lambda,
\gamma \Sigma^0$ reactions. Our main objective is to investigate whether the
data for these reactions can also be well 
described with the parameters fixed by the
SU(3) chiral symmetry and hence can provide us with an additional support to
the interpretation of the 
$\Lambda(1405)$ as a quasi-bound meson-baryon state with $S=-1$.

In section II, we present a  derivation of the approach \cite{oset} 
within a well-defined formulation of relativistic
quantum field theory. This will allow us to 
clearly establish the rules by which the approximations
were introduced in solving the problem. These rules then allow us to 
easily derive in section III 
the $K^- p \rightarrow \gamma Y$ amplitude that is consistent with
the model of Ref.\cite{oset}. The results and discussions are presented 
in section IV.

\section{Derivation of the unitary coupled channel method}
The approaches followed in Refs.\cite{kaiser,oset} were developed by 
implementing some physical 
considerations of chiral symmetry into a postulated coupled-channel
scattering equation.
To extend the model of Ref.\cite{oset} to investigate $K^- p \rightarrow
\gamma Y$ reactions, it is useful to establish its derivation from a 
well-defined formulation of relativistic quantum field theory.
 
Within the relativistic quantum field theory\cite{zuker}, the S-matrix for
the reactions we are considering can be defined as
\begin{eqnarray}
S_{ij} = \delta_{ij} - (2\pi)^4 i \delta^{(4)}(P_i - P_f) \hat{T}_{ij} \, ,
\end{eqnarray}
where $i,j$ denote either a meson-baryon($MB$) channel or a 
photon-hyperon($\gamma Y$) channel, and $P_i$ is the total
four-momentum of the system.  The scattering amplitude is defined by
\begin{eqnarray}
\hat{T}_{ij} = \frac{1}{(2\pi)^6}\frac{1}{\sqrt{2\omega _i}}
\sqrt{\frac{M_i}{E_i}}
T_{ij}\sqrt{\frac{M_j}{E_j}}\frac{1}{\sqrt{2\omega_j}} \, ,
\end{eqnarray}
where $M_i$ and $E_i$ denote respectively the mass and energy of the baryon
, $\omega_i$ the energy of the meson. Here we have defined
\begin{eqnarray}
T_{ij} = \bar{u}_i t _{ij} u_{j} \, ,
\end{eqnarray}
where $u_i$ is the Dirac spinor and the invariant amplitude $t_{ij}$
is defined by a Bethe-Salpeter equation(see, for example,
the derivation in Ref.\cite{zuker}). In momentum-space, it takes the 
following form
\begin{eqnarray}
t_{ij}(k_i,k_j; P) &=& I_{ij}(k_i,k_j;P) \nonumber \\
     & & + i \sum_{l} \int \frac{d^4k}{(2\pi)^4} 
 I_{il}(k_i,k;P)\frac{1}{(\not\!P-\not\!k)-M_l+ i \epsilon}
\frac{1}{k^2-\mu_l^2 + i \epsilon}t_{lj}(k,k_j;P) \,.
\end{eqnarray}
where $k$'s are the meson momenta, $P$ is the total momentum of the system,
$M_l$ and $\mu_l$ are respectively the masses of the baryon and meson
in the intermediate channel $l$. 
In the rest of the paper,
we present formulae in the center of mass system with $P=(\sqrt{s}, \vec{0})$.

The driving term in Eq.(4) is the sum of all two-particle 
irreducible amplitudes that
can be generated from the chosen SU(3) effective chiral lagrangian by using 
the standard perturbation theory.
To lowest order these amplitudes are
\begin{eqnarray}
I_{ij}(k_i,k_j) = \frac{-C_{ij}}{4 f^2}(\not\!k_i+\not\!k_j) \, ,
\end{eqnarray}
where $C_{ij}$ are SU(3) coupling constants which can be found in Ref.
\cite{oset}, and
$f$ is the pion decay constant.

The scattering equation employed in \cite{oset} can now be obtained from
Eq.(4) by using a simplification: only the positive energy component
of the Fermion propagator is kept.
Explicitly, the rule is to use the following substitution in evaluating
any loop-integration:
\begin{eqnarray}
\frac{1}{\not\!p-M + i\epsilon} &\rightarrow& \frac{M}{E(p)}
\frac{u_{\vec{p}}\bar{u}_{\vec{p}}}{p^0 - E(p) +i\epsilon} \, .
\end{eqnarray}
Substituting Eq.(6) into Eq.(4) and integrating out the time component $k^0$,
it is straightforward to use the definitions Eqs.(3) and (5) to obtain
\begin{eqnarray}
T_{ij}(k_i,k_j,\sqrt{s}) &=& V_{ij}(k_i,k_j) \nonumber \\ 
   & & +\sum_{l}\int \frac{d\vec{k}}{(2\pi)^3}\frac{M_l}{E_l(k)}
\frac{1}{2\omega_l(k)}\frac{V_{il}(k_i,k)T_{lj}(k,k_j,\sqrt{s})}
{\sqrt{s}-E_l(k)-\omega_l(k)+i\epsilon } \, ,
\end{eqnarray}
where 
\begin{eqnarray}
V_{ij}(k_i,k_j)=\bar{u}_{-\vec{k}_i}\frac{-C_{ij}}{4f^2}(\not\!k_i
+\not\!k_j) u_{-\vec{k}_j} \, .
\end{eqnarray}
In the near threshold region, one can use the heavy-baryon approximation to
reduce the potential into a spin-independent s-wave interaction
\begin{eqnarray}
V_{ij}(k',k) \rightarrow \frac{-C_{ij}}{4f^2}(k_0'+k_0)
\end{eqnarray}

With the definitions Eqs.(1)-(3) for the S-matrix and the s-wave 
spin-independent interaction
Eq.(9), it is straightforward to obtain the following expression of
the total cross section
\begin{eqnarray}
\sigma_{j\rightarrow i}(\sqrt{s}) = \frac{1}{4\pi}\frac{1}{s}\frac{k_i}{k_j}
M_iM_j \bar{\sum} \vert T_{ij}(\sqrt{s}) \vert ^2
\end{eqnarray}
where $\bar{\sum}$ stands for the sum
and average over the final and inital spin
indices.
Eqs.(7), (9), and (10) define precisely the model used in \cite{oset}.

In Ref.\cite{oset} Eq.(7) is solved by evaluating the potential $V_{ij}$ 
on-shell and hence it factors out of the integral. It was argued that 
the off-shell
effects are in the renormalization of coupling constants and can 
be neglected in the calculation using physical masses and coupling
constants. Eq.(7) then becomes a simple algebraic equation which can be solved
easily. A cutoff $q_{max} = 630 $ MeV is found to be needed to
regularize the integration in Eq.(7) and give a good description of
all of the existing $S=-1$ meson-nucleon reactions near threshold.
This factorization technique is similar to that
first introduced\cite{oller} in the
study of meson-meson scattering using chiral lagrangians, where the off-shell
contributions can be absorbed into the renormalization of masses and coupling
constants.

\section{The $K^- p \rightarrow \gamma Y$ amplitude}

We follow here in the $S=-1$ sector similar steps as done in 
\cite{saghai,kaiserbis} in the S=0 sector, but starting from a relativistic 
formulation of the problem.
To study the $K^- p \rightarrow \gamma Y$ reaction 
with $Y=\Lambda, \Sigma^0$, we return to
the Bethe-Salpeter Equation (4), setting $i=\gamma Y$ and $j= K^- p$.
To the leading order in the
electromagnetic coupling constant $e$, we have
\begin{eqnarray}
t_{\gamma Y,K^-p}(q, k'; P) &=& A_{\gamma Y,K^-p}(q, k';P) \nonumber \\
     & & + i \sum_{MB} \int \frac{d^4k}{(2\pi)^4} 
\, [A_{\gamma Y,MB}(q,k;P)\frac{1}{(\not\!P-\not\!k)-M_B+ i \epsilon} \nonumber \\
& & \times \frac{1}{k^2-\mu_M^2+ i \epsilon}t_{MB,K^-p}(k,k';P)  \, ]\, ,
\end{eqnarray}
where $MB$ denotes the allowed intermediate meson-baryon states. In general,
we need to include $MB = K^- p, \pi^- \Sigma^+, \pi^+ \Sigma^-, \bar{K}^0 n,
\pi^0 \Lambda, \pi^0 \Sigma^0, \eta\Lambda, \eta\Sigma^0, K^+ \Xi^-, 
K^0 \Xi^0$. Eq.(11) is illustrated in Fig.1. 

The photoproduction mechanism is described by the amplitude
$A_{\gamma Y, MB}$ in Eq.(11). 
Within the SU(3) effective chiral lagrangian including
the minimum electromagnetic coupling,
it has the form of the standard pseudovector 
Born term, as illustrated in Fig.2. 
Explicitly, we have
\begin{eqnarray}
A_{\gamma Y,MB} = A^{(a)}_{\gamma Y,MB}+A^{(b)}_{\gamma Y,MB}
+A^{(c)}_{\gamma Y,MB}+A^{(d)}_{\gamma Y, MB}  
\end{eqnarray}
with
\begin{eqnarray}
A^{(a)}_{\gamma Y,MB}(q,k)&= & i\sum_{Y'}\,[\frac{-C_{Y',MB}}{2f}\, ]
\epsilon_{\mu}F^{\mu}_{\gamma Y Y'}
\frac{1}{(\not\!q+\not\!p)-M_{Y'}+i\epsilon}
\gamma_5\not\!k  \, ,\\
A^{(b)}_{\gamma Y,MB}(q,k)&= & i\sum_{B'}\,[\frac{-C_{Y,MB'}}{2f}\, ]
\gamma_5\not\!k
\frac{1}{(\not\!p'-\not\!q)-M_{B'}+i\epsilon}
\epsilon_{\mu}F^{\mu}_{\gamma B'B} \, , \\
A^{(c)}_{\gamma Y,MB}(q,k)&=& i\,[\frac{-C_{Y,MB}}{2f}\, ]e_M
\frac{\gamma_5(\not\!k-\not\!q)\epsilon\cdot(2k-q)}
{(k-q)^2-\mu_M^2 +i \epsilon} \, , \\
A^{(d)}_{\gamma Y, MB}(q,k)&=& i[\frac{-C_{Y,MB}}{2f}\,] e_M 
\, [-\gamma_5\not\!\epsilon \, ] \, ,
\end{eqnarray}
where $e_M$ is the charge of the meson
$M$. The SU(3) coupling constants $C_{B',MB}$ for
the $MB \leftrightarrow B'$ transition are defined by \cite{pich}
\begin{eqnarray}
C_{B',MB}= X_{B',MB}(D+F) + Z_{B', MB} (D-F) \,
\end{eqnarray}
with\cite{pich} $D+F = g_A = 1.257$ and $D-F= 0.33$. The values of X's and Z's
needed for our calculations can be easily evaluated using the chiral
lagrangians of Ref.\cite{pich,bernard} and will be given later.
The photon-baryon-baryon vertices
are defined by 
\begin{eqnarray}
\epsilon_\mu F^{\mu}_{\gamma B B'} = e_B \delta_{BB'}\not\!\epsilon 
-\frac{\kappa_{BB'}}{4 M_p}
(\not\!\epsilon\not\!q - \not\!q\not\!\epsilon) \, ,
\end{eqnarray}
where $e_B$ is the charge of the baryon $B$ and $\kappa_{BB'}$ is the 
anamalous magnetic transition moment. The normalizations were chosen
such that $\kappa_{\gamma p p }= \kappa_{proton} = 1.79$ for the proton.

We now apply the rule of Eq.(6) to evaluate the integration in Eq.(11).
The derivation is straightforward. Here we only note that the 
integration over the meson-exchange term (Fig.2c) can have 
two meson pole contributions, and replacing 
the baryon propagators in Eqs.(13)-(14) by the projected propagator of Eq.(6) 
makes our photoproduction matrix elements different from the usual
Born terms used in Refs.\cite{workman,saghai}. 
In addition, we neglect terms of order of $(k/M_B)^2$ or $k\omega/M_B^2$
and higher, typical of the heavy baryon approximation. These 
approximations are required for
consistency with the construction of the coupled-channel
potential, Eq.(9), within the approach of \cite{oset}.
Accordingly, the strong amplitude $t_{MB, K^- p}$ in Eq.(11) is evaluated
with the on-shell momenta and factors out
of the integration in the derivation. As we show in the
appendix, this factorization comes 
from the fact that the off-shell corrections can be absorbed in the 
charge renormalization. This line of argumentation is based on the work 
\cite{gamma} for the $\gamma\gamma\rightarrow$meson-meson reaction, where it 
was first used.

With the above steps and the definition of the transition amplitude given in 
Eq.(3), we arrive at the following  
expression in the center of mass frame $P=(\sqrt{s},\vec{0})$ 
\begin{eqnarray}
T_{\gamma Y, K^- p}(q,k') &=& Q_{\gamma Y,K^- p}(q,k') 
+[QGT]_{\gamma Y,K^- p}(q,k')+ \Delta_{\gamma Y,K^- p}(q,k') \, .
\end{eqnarray}
The second term of the above equation 
has a familiar form for describing the initial state interactions
\begin{eqnarray}
[QGT]_{\gamma Y,K^-p}(q,k') &=& \sum_{MB}  
\,\{ [\int \frac{d\vec{k}}{(2\pi)^3}\frac{M_B}{E_B(k)}\frac{1}{2\omega_M(k)}
\frac{Q_{\gamma Y,MB}(q,k)}{\sqrt{s}-E_B(k)-\omega_M(k)
+i\epsilon}] \nonumber \\
& & \, \, \times T_{MB, K^- p}(k_{MB},k',\sqrt{s}) \} \, ,
\end{eqnarray}
where $k_{MB}$ is the on-shell momentum for the channel $MB$. 
The third term in Eq.(19) is due to the second pole of
meson-exchange term Fig.2c. It has the following form
\begin{eqnarray}
\Delta_{\gamma Y,K^- p}(q,k') &= i &\sum_{MB}\, \, \{ \, (
\int \frac{d\vec{k}}{(2\pi)^3} \, \frac{M_B}{E_B(k)}
\, [\frac{1}{2\omega_M(\vec{k}-\vec{q})} \, ]
\frac{1}{\sqrt{s}-E_B(k)-q^0 -\omega(\vec{k}-\vec{q})+i\epsilon} \nonumber \\
& & \, \, \times\,[\frac{-1}{2f}C_{Y,MB} \, ]e_M
\frac{-2\, [\vec{k}\cdot\vec{\epsilon}]
\,[\vec{\sigma}\cdot(\vec{k}-\vec{q})\,]}
{(q^0+\omega_M(\vec{q}-\vec{k}))^2-\omega_M^2(k)}  \, )  \nonumber \\
& & \, \, \times T_{MB,K^-p}(k_{MB},k) \, \} \, .
\end{eqnarray}

Keeping only the s-wave meson-baryon states and employing
the heavy-baryon approximation described above, the 
photoproduction amplitude in Eqs.(19) and (20) takes the following
form
\begin{eqnarray}
Q_{\gamma Y,MB}(q,k)= i\, [ \,\vec{\sigma}\cdot\vec{\epsilon}\, ] 
F_{\gamma Y,MB}(k,q) \, ,
\end{eqnarray}
where
\begin{eqnarray}
F_{\gamma Y,MB}(k,q)= - e_M 
\,[ \frac{-1}{2f}C_{Y,MB}\, ] 
\, [ 1 - \frac{\omega_M(k)}{2q}+\frac{\mu^2_M}{4qk}ln\frac{\omega_M(k)+k}
{\omega_M(k)-k} \, ] \, .
\end{eqnarray}
Note that the above expression is only from the meson-exchange term
(Fig.2c) and the contact term (Fig.2d). In the heavy-baryon approximation
employed here, one can show with some derivations that
the baryon pole term(Fig.2a) contributes only to the 
meson-baryon p-wave states, while
some s-wave contributions from the baryon-exchange term(Fig.2b) also
vanish at threshold.  
This simplicity is of course
due to our use of the baryon propagator shown in Eq.(6) to evaluate
the matrix elements of Eqs.(13) and (14), and the s-wave nature of
the meson-baryon channels within the approach followed in \cite{oset}.
Consequently, the total amplitude can be calculated by using Eqs.(19)-(22).
This also makes our predictions unambiguous since they do not depend on 
the less determined anomalous magnetic constants $\kappa_{B,B'}$
with $B, B' =$ hyperons in Eq.(18). 
Furthermore, the allowed
intermediate states will only be the charged particle 
channels $MB = K^-p, \pi^+ \Sigma^-, \pi^- \Sigma^+$ and $ K^+ \Xi$.
The coupling constants $C_{B',MB}$
needed for our calculation are defined in Eq.(17) with their
coefficients X's and Z's listed in Table I.

\section{Results and Discussions}

The calculations involve two parameters: 
the cutoff parameter $q_{max}=630 $ MeV for regularizing
the integrations in Eqs.(20) and (21) 
and the overall SU(3) coupling strength $f =1.15$ $F_\pi$($F_{\pi}=93$ MeV), 
approximate average between $f_\pi$ and $f_K$.
These two parameters were determined in Ref.\cite{oset} and hence  the
calculations based on Eqs.(19)-(22) do not have any adjustable parameters.

We have calculated the branching ratios defined by
\begin{eqnarray}
B_{K^- p \rightarrow \gamma Y} = \frac{\sigma_{K^- p \rightarrow \gamma Y}
(\sqrt{s}_{th})}
{\sigma_{K^- p \rightarrow all}(\sqrt{s}_{th})} \, ,
\end{eqnarray}
where $Y= \Lambda, \Sigma^0$ and the total cross sections are 
defined in Eq.(10). In the calculation of the denominator of Eq.(24)
all $S=-1$ meson-baryon channels within the approach of \cite{oset} 
are included.
The results presented
below are obtained by evaluating the above
expression at $\sqrt{s}_{th}\rightarrow \mu_{K^-} + M_{p}$.

Our results calculated with and without including the coupled-channel
effects are listed in Table II and compared with the data \cite{data}.
We see that with no initial meson-baryon interactions (first row in Table II),
the $\gamma \Sigma^0$ production is very weak and
the predicted ratio between two production rates is 
an order of magnitude larger than the data. This is in
agreement with the findings of Siegel and Saghai\cite{saghai}. When
the strong coupled-channel effects are included our predicted
ratio(second row of Table II) is close to the experimental value.
The predicted branching ratio for the $\gamma \Lambda$ production
is about 50 $\%$ larger than the experimental value, but it is within 
the experimental uncertainty for the $\gamma \Sigma^0$ production.

To understand the dynamical origins of our results listed in Table II, we 
examine the coupled-channel effects in some detail.
For this purpose it is sufficient to just consider 
the first two terms of Eq.(19). They can be
cast into a form employed by Siegel and Saghai\cite{saghai}, if we 
factor out the photoproduction amplitude $Q_{\gamma Y, MB}$ and
evaluate it at the on-shell momenta. With this approximation, 
we have( obvious variables are omitted here)
\begin{eqnarray}
T_{\gamma Y, K^-p} \sim
\sum_{MB} F_{\gamma Y,MB}[1+GT]_{MB,K^-p} \,.
\end{eqnarray} 
All quantites in the above equation are evaluated at on-shell momenta.
Obviously the term $[1+GT]_{MB,K^-p}$ measures the strength of the 
$K^- p \rightarrow MB$ transition. In Table III, we list these matrix elements as well
as the values of the on-shell photoproduction amplitudes 
$F_{\gamma Y, MB}$. We see that the total amplitude, defined by Eq.(25)
in this estimate, involves a
nontrival interplay between the strong transitions and electromagnetic 
transitions. In particular, the 
large enhancement of the $\gamma \Sigma^0$ production
by the coupled-channel effects can be understood from the
second column of Table III. We see that
$F_{\gamma \Sigma^0, \pi^+ \Sigma^-}$ and 
$F_{\gamma \Sigma^0, \pi^- \Sigma^+}$ are a factor of about 3 larger 
than $F_{\gamma \Sigma^0, K^- p}$, while their strong transition
strengths in column 3 are comparable in magnitude.  Consequently,
the branching ratio for $\gamma \Sigma^0$ is greatly enhanced when 
the $\pi$ channels are included.
This is verified in our exact calculations based on Eqs.(19)-(20), as
can be seen in the second row of Table IV. We see that
the  branching ratio for $\gamma \Sigma^0$ production
is increased from 0.14 to 1.28 when the $\pi^+$ and $\pi^-$ channels
are included. From column 1 of Table III, we also expect that the
effect of $\pi$ channels 
on the $\gamma \Lambda$ production is much less, mainly because the
photoproduction amplitudes for the first three channels are comparable 
and the contributions from $\pi^+$ and $\pi^-$ channels have opposite signs.
As shown in the first row of Table IV, the coupling with
the $\pi$ channels can lower the production rate only from 2.47 to 1.56. 
Neverthelesss, it is instrumental in  bringing our prediction 
closer to the experimental value $0.86\pm 0.16$.
We further notice that the reduction from 2.47 to 1.33 when the
$\pi^+ \Sigma^-$ channel is included (first row of Table IV) 
is largely due to the cancellation caused by the opposite signs of 
the imaginary parts of $[1+GT]$, as seen in the third column of Table III. 
Note that the relative signs between the contributions from 
different channels are determined not only by the meson charges($e_M$) but
also by the chiral SU(3) parameters listed in Table I. Thus,
the coupled-channel effects are crucial in testing the
chiral SU(3) symmetry. The effect due to the $K^+ \Xi^-$ channel is very 
weak because their strong transition is an order of magnitude smaller
than others, as can be seen in the third column of Table III.

We now turn to illustrating several important features of our approach.
The exact treatment of meson propagators in Fig.2c leads to a 
contribution($\Delta_{\gamma Y, K^-p}$ in Eq.(19))
from a second pole. By comparing the first two rows in Table V, we
see that this second pole term can change the $\gamma \Sigma^0$
branching ratio by about 40$\%$, but much less for $\gamma \Lambda$.
Consequently, the the predicted ratio becomes closer to the experimental value.

An important finding of Ref.\cite{oset} was that the coupling with the 
$\eta$ channels is essential in obtaining a good agreement with all of 
the existing $S=-1$ meson-baryon reaction data when using only the lowest 
order lagrangian as input. The influence of this coupling
on our predictions is also significant. This can be seen by
comparing the second and third rows
in Table V. We see that the predicted
branching ratio for $\gamma \Lambda$ production is increased by about
60 $\%$ if the $\eta$ channels are omitted in the calculation of the 
strong amplitudes $T_{MB,K^- p}$ appearing in Eq.(19). It is clear that 
including the
$\eta$ channels is also crucial in using this reaction to test
the chiral SU(3) symmetry. We note that the $\eta$ channels were
omitted in the model of Ref. \cite{saghai}. At the same time the
couplings in that work had to be substantially changed with 
respect to their SU(3) values in order to obtain a good fit to
the data. In retrospective we can say that what the explicit breaking 
of SU(3) symmetry does in reality is to
restore it after it has been broken by the omission of the
$\eta$ channels.

Finally, we note that the strong meson-baryon-baryon vertex in each of the 
photoproduction amplitudes should in principle have a form factor 
because hadrons are composite particles. To see how our results will 
be changed when this is taken into account, we perform
a calculation including a monopole form factor with typical cut off, 
$\Lambda_\pi$, values of 1.2 and 1  GeV. 
The introduction of this cut off reduces the ratio $R$ to values
in better agreement with the data. Changes of $\Lambda_\pi$ from
1 to 1.2 GeV introduce only corrections of the order of 10\%.
The results obtained with $\Lambda_\pi=1$ GeV, a value which is commonly
accepted, roughly agree with the data within experimental errors, 
which are of order of 20\%. If one compares with the central values of the 
experimental branching ratios, our results are on the upper edge 
of the $B_{K^- p \to \gamma \Lambda}$ ratio while those for
$B_{K^- p \to \gamma \Sigma^0}$ are on the lower edge. Looked at it
in the context that the coupled channels and unitarization have
reduced the ratio $R$ by a factor 14, differences of the order of
10--20 \% are not so significative. Yet, the fact that a better
agreement with the central values of the data is obtained in
\cite{saghai}, at the price of fitting parameters, indicates that
there is probably still room in the model used here for
moderate breakings of SU(3) beyond those implemented by the 
different masses of the particles.

In conclusion, we have extended the approach of \cite{oset} to
make predictions for the branching ratios for the $K^- p
\rightarrow \gamma \Lambda, \gamma \Sigma^0$ reactions near the 
threshold. All coupling constants are consistent with the chiral SU(3) 
symmetry. With only two parameters, which were fixed in the
study of $S=-1$ meson-baryon reactions, our predictions are close to the
data, in particular the ratio between two branching ratios.
In our approach, neither the meson-baryon nor the photoproduction 
mechanisms involve the explicit consideration of excited hyperon states 
since the $\Lambda(1405)$ resonance, which plays a key role in these reactions, 
is generated dynamically. 
Our results therefore strengthen the interpretation of the
$\Lambda(1405)$ as a quasi-bound meson-baryon system with $S=-1$, as already 
supported by the study of the strong interactions in \cite{kaiser,oset}. This 
does not exclude a seed of a more complicated intrinsic quark substructure, 
but the strong coupling to the meson baryon channels imposed by the 
unitarization reinforces the meson baryon component and allows this 
resonance to approximately qualify as a meson baryon quasibound state 
much as it happens for the resonances in the scalar meson-meson sector 
\cite{oller2,torn}

\bigskip
\bigskip
\bigskip

{\bf Acknowledgements}
\bigskip

One of the authors T.-S. H. Lee wishes to acknowledge the financial support
from the Programa C\'atedra of BBV and the warm hospitality of Professor E.
Oset of University of Valencia. J.A.O. wishes to acknowledge financial
help from the Generalitat Valenciana. We would like to acknowledge useful 
comments from W. Weise. This work is partially supported 
by DGICYT contract numbers PB95-1249 and PB96-0753, and also by US 
Department of Energy, Nuclear Physics Division, under contract No. 
W-31-109-ENG-38. The authors acknowledge partial support from the
EEC-TMR Program under contract No. CT98-0169.
\newpage
\appendix
\section{Renormalizability character of the off shell part 
of the strong amplitude.}
We start from Eq. (11) and take just one loop in the Bethe  Salpeter equation. 
This means that we substitute $t_{MB,K^- p}(k,k',P)$ by the $I_{ij}(k,k',P)$ 
function of Eq. (5). Let us take the contact term for the electromagnetic 
amplitude $A_{\gamma Y,MB}$ (the procedure and the conclusion for the other 
terms are the same). By taking the nonrelativistic reduction of the contact 
term and the positive energy part of the nucleon propagator, Eq. (6), as done 
in the evaluation of the meson nucleon strong amplitude in \cite{oset}, we 
obtain from the loop a contribution proportional to 
\begin{equation}
\label{aa}
i\int\frac{d^4k}{(2\pi)^4}\frac{M}{E_B(k)}\frac{1}{p^0+k'^0-k^0-E_B(\vec{k})
+i\epsilon}(k'^0+k^0)\frac{1}{k^2-\mu_M^2+i\epsilon}e\vec{\sigma}\vec{\epsilon}
\end{equation}

We now separate the strong amplitude, represented by the factor $k'^0+k^0$ into 
its on shell and off shell parts as

\begin{equation}
\label{ab}
k'^0+k^0=2 k'^0+(k^0-k'^0)
\end{equation}

The part of the integral coming from the term $2k'^0$ in the former 
equation corresponds to factorizing the strong amplitude with its on 
shell value, which is the procedure that we have followed. The 
contribution from the off shell part will be given by 

\begin{equation}
\label{ac}
i\int\frac{d^4k}{(2\pi)^4}\frac{M}{E_B(k)}\frac{1}{p^0+k'^0-k^0-E_B(\vec{k})
+i\epsilon}(k^0-k'^0)\frac{1}{k^2-\mu_M^2+i\epsilon}e\vec{\sigma}\vec{\epsilon}
\end{equation}
Here we follow the same steps that led to include the off shell contribution 
in the strong amplitude into a renormalization of the $f$ coupling in 
\cite{oset}. We use typical approximations of the heavy baryon formalism 
and set $M/E_B(\vec{k})=1$, $p^0-E_B(\vec{k})=0$. In this case the off 
shell term $(k^0-k'^0)$ and the nucleon propagator $(k'^0-k^0)^{-1}$ cancel and 
we get a contribution 

\begin{eqnarray}
\label{ad}
&&-i\int\frac{d^4k}{(2\pi)^4}\frac{1}{k^2-\mu_M^2+i\epsilon}e\vec{\sigma}\vec{\epsilon}
\nonumber\\
&=&-\int\frac{d^3k}{(2\pi)^3}\frac{1}{2w(\vec{k})}e\vec{\sigma}\vec{\epsilon}\sim 
q_{max}^2e\vec{\sigma}\vec{\epsilon}
\end{eqnarray}

Hence, we obtain a contribution proportional to the cut off momentum squared, 
independent of energy and with the same structure as the contact term in the 
Born amplitude, which 
goes into renormalizing the contact term and is taken into account when using 
this term with the physical electromagnetic coupling $e$.

The generalization to other terms of the electromagnetic amplitude and higher 
order loops proceeds analogously with the conclusion that only the on shell 
part of the strong amplitude must be used.

\narrowtext

\begin{table}
\caption{SU(3) coupling constants defined in Eq.(17)}
\label{TAB1} 
\begin{tabular}{|c|ccccc}
$X_{B'MB}$ &$MB=$ &$K^- p$ &$\pi^+ \Sigma ^- $ &$\pi^- \Sigma^+$ &$ K^+ \Xi ^- $\\
\hline
$B'=\Lambda$ &$- $ &$\frac{-2}{\sqrt{3}}$ &
$\frac{1}{\sqrt{3}}$ &$\frac{1}{\sqrt{3}}$
&$ \frac{1}{\sqrt{3}}$ \\
$B'=\Sigma^0$ &$- $ &$ 0$ &$1$ &$-1$ &$ 1$ \\
\end{tabular}
\begin{tabular}{|c|ccccc}
$Z_{B'MB}$ &$MB=$ &$ K^-p$ &$ \pi^+\Sigma^-$ &$ \pi^-\Sigma^+$ &$ K^+ 
\Xi^- $\\
\hline
$B'=\Lambda$ & $-$ &$\frac{1}{\sqrt{3}}$ 
&$\frac{1}{\sqrt{3}}$ &$\frac{1}{\sqrt{3}}$
 &$ \frac{-2}{\sqrt{3}}$ \\
$B'=\Sigma^0$ &$-$ &$ 1$ &$-1$ &$ 1$ &$ 0$ \\
\end{tabular} 
\bigskip
\bigskip
\bigskip
\caption{Comparisions of the predicted $K^- p\rightarrow \gamma\Lambda,
\gamma\Sigma^0$ branching ratios defined in 
Eq.(24)(in unit of $10^{-3}$) with the data[20]. The
amplitudes are defined in Eqs.(19)-(22).}
\label{TAB2}
\begin{tabular}{rccc}
\hspace*{20mm} $Amplitude$ &$B_{K^-p\rightarrow \gamma\Lambda}$&
$B_{K^-p\rightarrow \gamma\Sigma^0}$&$R=B_{K^-p\rightarrow\gamma 
\Lambda}/B_{K^-p\rightarrow \gamma \Sigma^0}$\\
\hline
   $Q$ &1.12 &0.073 &16.4 \\
   $Q + GQT +\Delta  $ &1.58 &1.33& 1.19   \\
    Data \protect\cite{data} &$0.86\pm0.16$&$1.44\pm0.31$&0.4 - 0.9   
\end{tabular}
\bigskip
\bigskip
\bigskip
\caption{The quantities defined in Eq.(25)
are compared. The coefficient 
$F_{\gamma Y,MB}$ is defined in Eq.(23) and is in unit of 
$10^{-2}$/MeV.}
\label{TAB3}
\begin{tabular}{rccc}
$MB$&$F_{\gamma\Lambda,MB}$&$F_{\gamma\Sigma^0,MB}$&$[1+GT]_{MB,K^-p}$\\
\hline
$K^-p$ & 0.588& -0.159&-0.68+i1.63 \\
$ \pi^+ \Sigma^- $ &0.431&0.430&-0.66-i1.01\\
$\pi^-\Sigma^+$&-0.431&0.430&-0.61-i0.40\\
$K^+\Xi^-$&0.157&0.589&-0.087-i0.002
\end{tabular}
\bigskip
\bigskip
\vspace{2cm}
\caption{Branching ratios(in unit of $10^{-3}$) predicted from the 
calculations with different numbers of coupled-channels included. 
Here we define
$Ratio=B_{K^-p\rightarrow\gamma
\Lambda}/B_{K^-p\rightarrow \gamma \Sigma^0}$}
\label{TAB4}
\begin{tabular}{rccccc}
&&&&Channels included&\\
&0&$K^-p$&$K^-p + \pi^+\Sigma^-$&$K^-p + \pi^+\Sigma^- + \pi^-\Sigma^+$&$
K^-p + \pi^+\Sigma^- + \pi^-\Sigma^+ + K^+ \Sigma^-$\\
\hline
$B_{K^-p\rightarrow\gamma\Lambda}$&1.12&2.47&1.33&1.56&1.58\\
$B_{K^-p\rightarrow \gamma\Sigma^0}$&0.073&0.14&0.74&1.28&1.33\\
$Ratio$&16.4&17.5&1.81&1.22&1.19
\end{tabular}
\bigskip
\bigskip
\bigskip
\caption{Same as Table II. The subindex $no-\eta$ indicates that
the strong amplitude $T_{K^- p, MB}$ is calculated with the $\eta $ channels
turned off. The subindex $with-\Lambda_{\pi}$  indicates that a monopole
form factor with cutoff $\Lambda_{\pi}=1$ GeV (1.2 GeV in brackets)
is included 
in the meson-baryon-baryon vertices of photoproduction
amplitudes.}
\label{TAB5}
\begin{tabular}{rccc}
\hspace*{20mm} $Amplitude$&$B_{K^-p\rightarrow \gamma\Lambda}$&
$B_{K^-p\rightarrow \gamma\Sigma^0}$&$R=B_{K^-p\rightarrow\gamma 
\Lambda}/B_{K^-p\rightarrow \gamma \Sigma^0}$\\
\hline
   $Q + QGT  $ &1.31 & 0.95&1.38 \\
   $Q + QGT + \Delta $ &1.58 &1.33 &1.19    \\
   $\, [Q + QGT+\Delta \,]_{no-\eta}  $ &2.47 &1.27&1.94 \\
   $\, [ Q + QGT+\Delta \, ]_{with-\Lambda_\pi}$ &1.10 (1.22) & 1.05 (1.13) 
       & 1.04 (1.08) \\
Data \protect\cite{data} &$0.86\pm0.16$ & $1.44\pm0.31$&$0.4 - 0.9$
\end{tabular}
\end{table}
\newpage
\begin{figure}
\caption{Graphical representation of Eq.(4)}
\end{figure}

\begin{figure}
\caption{Photoproduction mechanisms of Eq.(13)-(16)}
\end{figure}


\newpage

\begin{thebibliography}{99}
\bibitem{jones}
M. Jones, R.H. Dalitz and R.R. Horgan, Nucl. Phys. B{\bf 129} (1977) 45.

\bibitem{darewych}
J.D. Darewych, R. Koniuk and N. Isgur, Phys. Rev. D{\bf 32} (1985) 1765

\bibitem{veit}
E.A. Veit, B.K. Jennings, A.W. Thomas and R.C. Barrett, 
Phys. Rev. D{\bf 31} (1985) 1033.

\bibitem{zhong}
Y.S. Zhong, Q.W. Thomas, B.K. Jennings and R.C. Barrett,
Phys. Rev. D{\bf 38} (1988) 837.

\bibitem{kaxiras}
E. Kaxiras, E.J. Moniz and M. Soyeur, Phys. Rev. D{\bf 32} (1985) 695.

\bibitem{he}
G. He and R.H. Landau, Phys. Rev. C{\bf 48} (1993) 3047.

\bibitem{arima}
A. Arima, S. Matsui and K. Shimizu, Phys. Rev. C{\bf 49} (1994) 2831.

\bibitem{workman}
R.L. Workman and Harold W. Fearing, Phys. Rev. D{\bf 37} (1988) 3117;
J. Lowe, Nuovo Cimento, A{\bf 102} (1989) 16, and other
references therein. 

\bibitem{saghai}
Peter B. Siegel and Bijan Saghai, Phys. Rev. C{\bf 52} (1995) 392.

\bibitem{gasser}
J. Gasser and H. Leutwyler, Nucl. Phys. B{\bf 250} (1985) 465

\bibitem{meissner}
U.G. Meissner, Rep. Prog. Phys. {\bf 56} (1993) 903

\bibitem{pich}
A. Pich, Rep. Prog. Phys. {\bf 58} (1995) 563.

\bibitem{ecker}
G. Ecker, Prog. Part. Nucl. Phys. {\bf 35} (1995) 1.

\bibitem{bernard}
V. Bernard, N. Kaiser and U.G. Meissner, Int. J. Mod. Phys. {\bf E4} (1995) 193.

\bibitem{kaiser}
N. Kaiser, R. Siegel and W. Weise, Nucl. Phys. A{\bf 594} (1995) 325.

\bibitem{dalitz} R. H. Dalitz and S.F. Tuan, Ann. Phys. (N.Y.) {\bf 10} (1960)
307; R.H. Dalitz, T.-C. Wong and G. Rajasekaran, Phys. Rev. {\bf 153}
(1967) 1617.

\bibitem{martin} A.D. Martin, Nucl. Phys. B{\bf 179} (1981) 33.

\bibitem{kaiserbis}
N. Kaiser, T. Wass and W. Weise, Nucl. Phys. A{\bf612} (1997) 297.

\bibitem{oset} E. Oset and A. Ramos, Nucl. Phys. A{\bf 635} (1998) 99.

\bibitem{oller}
J.A. Oller and E. Oset, Nucl. Phys., A{\bf 620} (1997) 438

\bibitem{ramonet}
J.A. Oller, and J.R. Pelaez, Phys. Rev. Lett.{\bf80} (1998) 3452, 
in print, hep-ph/9803242

\bibitem{assum}
A. Parre\~no, A. Ramos and E. Oset, in preparation.

\bibitem{zuker}
C. Itzykson and J.B. Zuber, {\bf Quantum Field Theory} (McGraw Hill, 1980)

\bibitem{gamma} 
J.A. Oller and E. Oset, Nucl. Phys. A{\bf629} (1998) 739.

\bibitem{data}
D.A. Whitehouse et al., Phys. Rev. Lett {\bf 63} (1989) 1352.

\bibitem{oller2}
J.A. Oller and E. Oset, University of Valencia preprint. 

\bibitem{torn}
N.A. Tornqvist, Phys. Rev. Lett. {\bf49} (1982) 624;
K.L.Au, D. Morgan and M.R. Pennington, Phys. Rev. D{\bf35} (1987) 1633.  

\end{thebibliography}
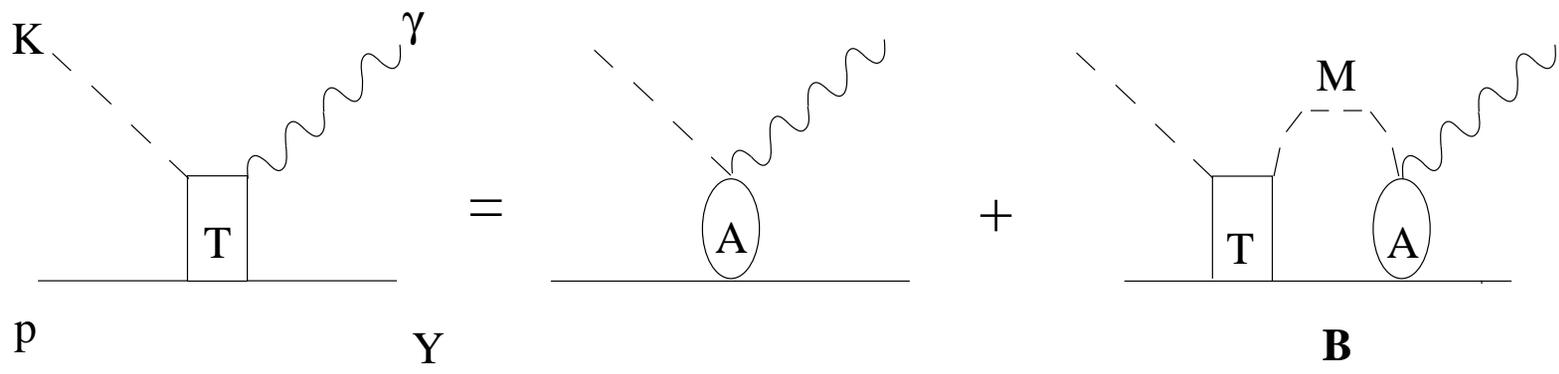
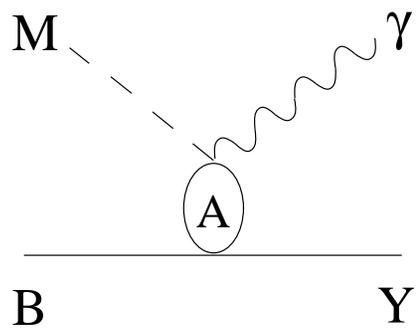
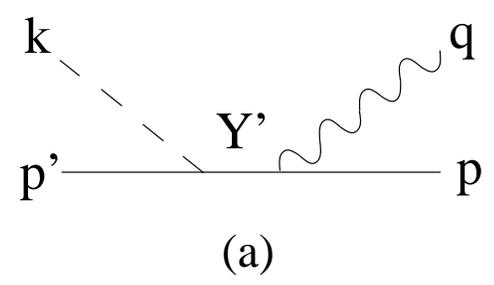
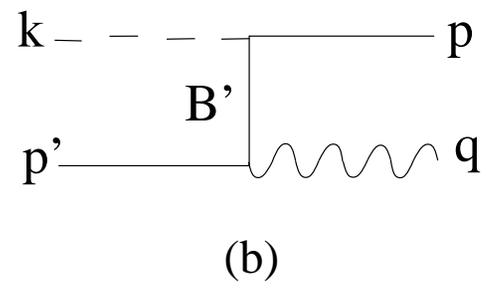
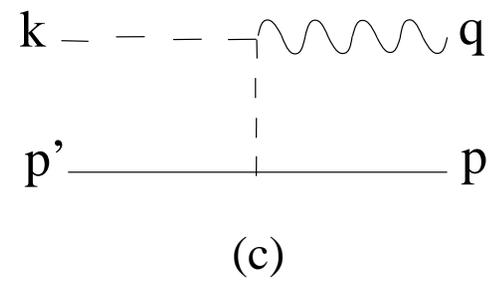
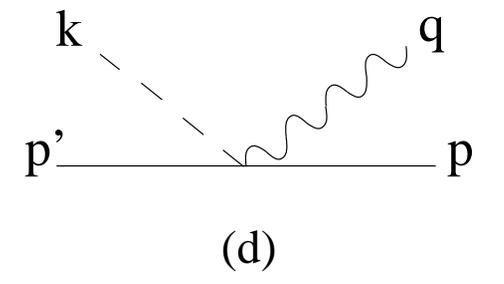
\end{document}